\documentclass[conference]{IEEEtran}

\ifCLASSINFOpdf
\else
\fi
\usepackage{graphicx}
\usepackage{slashbox}
\usepackage{multirow}
\usepackage{amsmath}

\begin{document}

%
\title{The Implications from Benchmarking Three Big Data Systems}


\author{
Jing Quan$^{1}\ \ \ $Yingjie Shi$^{2}\ \ \ $Ming Zhao$^{3}\ \ \ $Wei Yang$^{4}$\\
\\
\begin{tabular}{c}
$^1$School of Software Engineering, University of Science and Technology of China, Hefei, China \\
$^2$Institute of Computing Technology, Chinese Academy of Sciences, Beijing, China\\
$^3$School of Computing and Information Sciences, Florida International University, Florida, USA\\
$^4$School of Computer Science and Technology, University of Science and Technology of China, Hefei, China\\
\end{tabular}
\\
\\
\begin{tabular}{cccc}
sea11424@mail.ustc.edu.cn &shiyingjie@ict.ac.cn &ming@cs.fiu.edu &qubit@ustc.edu.cn\\
\end{tabular}
\\
}

\maketitle

\begin{abstract}
Along with today's data explosion and application diversification,
a variety of hardware platforms for big data are emerging, attracting interests from both industry and academia.
The existing hardware platforms represent a wide range of implementation approaches,
and different hardware platforms have different strengths.
In this paper, we conduct comprehensive evaluations on three representative big data systems: Intel Xeon, Atom (low power processors), and many-core Tilera using BigDataBench \--- a big data benchmark suite.
Then we explore the relative performance of the three implementation approaches by running BigDataBench,
and provide strong guidance for the big data systems construction.
Through our experiments,
we have inferred that a big data system based on specific hardware has different performance
in the context of different applications and data volumes.
When we construct a system,
we should take into account not only the performance or energy consumption of the pure hardware,
but also the characteristics of applications running on them.
Data scale, application type and complexity should be considered comprehensively
when researchers or architects plan to choose fundamental components for their big data systems.

\end{abstract}

\IEEEpeerreviewmaketitle

\section{Introduction}
We can hardly stop the world creating data,
and it is not easy to estimate how much data is produced every day.
According to \textit{IDC} (International Data Corporation),
the total amount of global data is expected to grow to 2.7 zettabytes during 2012.
This is an increase of 48\% from 2011 \cite{shirer2011idc}.
The rapid growth of data volumes brings big challenges to big data systems,
and makes lots of existing solutions no longer applicable.
In response to this trend,
both academia and industry are looking for innovative ways to deal with large-scale data.
The distributed cluster system is a kind of solutions which is adopted by many organizations.
However, data volumes are growing to the point
where companies are being forced to strengthen their infrastructure,
for example, by adding more nodes or replacing with more powerful hardware.
The infrastructure costs and energy consumption skyrocket along this way.
Then, how to select proper hardware for big data systems,
and how to find a balance between processing capacity and energy costs
become an urgent issue.

Because of architectural difference, configuration variance, and other factors,
different hardware have their own processing specialities,
which would make them show diverse behaviors when dealing with different types of applications.
Architects are searching for a approach to take good use of different kinds of hardware resources to construct a balanced system.
And finding appropriate fundamental hardware components for the workloads counts for much more in constructing big data system.

In this paper,
we select a big data benchmark suite \--- \textit{BigDataBench} \cite{gao2013bigdatabench}\cite{BigDataBench} to evaluate three big data systems consisted of \textit{Xeon}, \textit{Atom}, and \textit{Tilera} respectively.
We choose them based on their popularity in industry and academia.
Xeon and Atom are widely used in many academic institutions and internet service providers,
and Tilera is a kind of burgeoning processor which has been applied to Facebook recent years \cite{berezecki2011many}.
What's more, these three platforms have great differences on their architecture.
The processors we use are Xeon E5310, Atom D510 and Tilera TilePro36.
Xeon E5310 adopts Core Micro-Architecture, supports \textit{OoO} (Out-of-Order) execution,
has strong processing capacity \cite{xeonwhitepaper}\cite{doweck2006inside}.
Atom D510 bases on Intel Pine Trail, doesn't support \textit{OoO}, possesses low power \cite{atomwhitepaper}.
Tilera is a many-core processor, it integrates 36 tiles (core is named as tile in Tilera) on one processor, adopts iMesh technology \cite{manycore} to handle core communication, which is totally different from Intel architecture.

Through experimental results collected,
we find Xeon has better processing capacity while has larger energy consumption in general.
The level of its advantage on processing capacity depends on the application type and data volume.
For I/O intensive application like Sort, the processing capacity gap between Xeon and the other two will shorten with the increasing of data volume.
For CPU intensive application like Grep, Wordcount and so on, the gap will keep stable when data volume approaches a level.
Atom and Tilera have low processing capacity relatively, and have low energy consumption on some light-weight applications such as Sort, Wordcount, and Grep.
When dealing with complex application like Naive Bayes or SVM, neither of them show obvious preponderance on energy performance.
More, Tilera shows better performance when processing I/O intensive application and is inadequate to process CPU intensive application.
So the merits of a system couldn't not be determined by hardware equipment simply,
it is also relative to the application type and the data volume running on it.
When architects choose fundamental components for their systems,
they can not only consider the performance or energy consumption of hardware itself,
but also need to regard the application type, data volume and the complexity of the application the system will run for.

The remainder of the paper is organized as follows.
Section 2 introduces our evaluation methodology, and elaborates the basic analysis on hardware platforms and workloads.
Section 3 discusses the experimental results based on three big data systems, and carries on the
comparative analysis to the processing results.
Section 4 concludes the paper.
\section{Evaluation Methodology}
In this section, we will give brief information about the hardware platforms we evaluate,
the Benchmarks we use, and the metrics we choose.
\subsection{Experimental Platforms Analysis}
As mentioned before, we choose three typical hardware platforms and evaluate their performance based on big data applications. The basic information of these three platforms are shown in TABLE \ref{xeon}\~{}TABLE \ref{tilera}.  We construct different-sized systems using these hardware platforms, and deploy Hadoop on them.
Hadoop is an open-source software for reliable, scalable, distributed computing \cite{hadoopweb},
and it is adopted in big data system by many companies,
such as Facebook \cite{berezecki2011many}, IBM \cite{ibmhadoop}, NREL \cite{hpgov}, etc.
The Hadoop version we used is Hadoop-1.0.2, and the setting refers to the guidance on Hadoop official web site.
The three platforms own different implementation architectures and characteristics, we make brief comparisons among them before showing the experiment result analysis.
\subsubsection{Xeon}
The hardware configuration of Xeon is showed in TABLE \ref{xeon}.
In the field of industry, Xeon usually is used to deal with resource-intensive applications, like JAVA or PHP.
Xeon E5310 is based on Intel Core Microarchitecture \cite{xeonwhitepaper}\cite{doweck2006inside}.
The processors has 4 cores per CPU with 1.6GHz basic frequency,
processes multiple arithmetic units and supports many instruction sets such as MMX, SSE and etc.
This model does not employ Hyper-Threading technology which means it has one hardware thread per core.
E5310 actually encapsulates two Woodcrest \cite{woodcrest} cores into one processor, so its L2 cache is 4MB*2, that is means the cache can't be shared by all 4 cores.
It adopts the Intel SpeedStep technology to adjustment the power according to the demand.
TDP (Thermal Design Power) of the chip is 80W.
\subsubsection{Atom}
Atom is often used in handling lightweight tasks, like web, Apache,
and some real-time applications.
All of these tasks split a problem into many pieces in order to leverage each core's processing power.
The configuration information is showed in TABLE \ref{atom}.
Atom D510 is based on Intel Pine Trail architecture \cite{atomwhitepaper}\cite{trail}, and its architecture is different from mainstream ones.
Atom uses the In-Order execution to reduce power consumption.
However, this is bound to affect processing performance.
Atom core introduces the hyper-threading (HT) technology, one core is equipped with dual-threaded executions.
The adoption of HT technology improves parallelism and make up for the lack of powerful execution architecture.
Its L2 cache is 512KB*2 (512KB per core).
The TDP of Atom is 13W.
\subsubsection{Tilera}
Tilera is a Many-Core processor for cloud applications such as memcached, media, and Hadoop \cite{manycore}.
In order for the tiles (cores) to communicate with each other and to I/O device,
the Tile Processor Architecture provides a communication fabric called the \textit{iMesh} which differ from \textit{BUS} in Intel Architecture.
Tilera also aims at lower power efficiency to adapt to massive-scale clusters.
The core number of Tilera series ranges from 16 to 100.
The information of Tilera we used is showed in TABLE \ref{tilera}.
TilePro36 integrates 36 tiles (cores) on one processor, and each tile is equipped with 16KB instruction cache, 8KB data cache, and 64KB L2 cache.
More, it provides virtual L3 cache. The L2 cache are shared by all tiles, then the L2 cache of one tile can be treated as the L3 cache of other tiles. Tilera offers relevant read policies.
Tilera does not support floating point operation,
and its tiles (cores) can be closed if need.
The TDP of TilePro36 is 16W.

In order to make the performance comparison fair and reasonable, we guarantee at least one dimension of two systems to be the same.
\begin{itemize}
\item During the comparison between Xeon and Atom, we make them have the same hardware thread number.
\item During the comparison between Xeon and Tilera, we make them have the same core number.
\end{itemize}

We construct the system on Xeon and Atom with one master and seven slaves.
There are 4 cores in a Xeon processor with one hardware thread per core,
and 2 cores in an Atom processor with two hardware threads per core.
So, the Xeon and Atom system have the same hardware thread number.
The Tilera processor we used has 36 tiles,
we closed 8 cores of it to make this system have same core number as Xeon.
Tilera will leave tiles on standby if the workload doesn't need to use full resource, so closing serval tiles won't have big influence on the function of other tiles.

\begin{table}
\caption{the basic configuration of Xeon}\label{xeon} \center
\begin{tabular}{|c|c|}
  \hline
  CPU Type & Intel \textregistered Xeon E5310\\ \hline
  CPU Core & 4 cores @ 1.6GHz \\ \hline
  L1 I/D Cache& 32KB  \\ \hline
  L2 Cache& 4096KB \\ \hline
\end{tabular}
\end{table}
\begin{table}
\caption{The basic configuration of Atom}\label{atom} \center
\begin{tabular}{|c|c|}
  \hline
  CPU Type & Intel \textregistered Atom D510\\ \hline
  CPU Core & 2 cores @ 1.66GHz \\ \hline
  L1 I/D Cache& 24KB \\ \hline
  L2 Cache & 512KB \\ \hline
\end{tabular}
\end{table}
\begin{table}
\caption{The basic configuration of Tilera}\label{tilera} \center
\begin{tabular}{|c|c|}
  \hline
  CPU Type & Tilera \textregistered TilePro36\\ \hline
  CPU Core& 36 cores @ 500MHz \\ \hline
  L1 I/D Cache&16KB/8KB \\ \hline
  L2 Cache  & 64KB \\ \hline
\end{tabular}
\end{table}

\subsection{Benchmark Selection}
We attempt to find some underlying relations between workloads and different big data systems in this paper,
especially the various impacts from different application type, and data volumes.
So, we need a benchmark, which can provide typical applications, and can offer data sets of any volume.
BigDataBench meets these all.

BigDataBench is a big data benchmark suite from web search engines.
It provides six kinds of applications which are typically employed in search engines,
including Sort, Grep, Wordcount, Naive Bayes, SVM and search \cite{gao2013bigdatabench}\cite{BigDataBench}.
And then, applications like Sort, Grep, Wordcount, have disparate computation complexity compared with Naive Bayes and SVM.
This will help us obtain more comprehensive results.
BigDataBench also provides a data generation tool to overcome difficulties of obtaining real big data.
This tool generates big data based on small-scale data while preserving the key characteristics of real data.
This can make our experimental results more reasonable.
We choose five out of six applications from BigDataBench.
For each application, the core operations are different.
From paper \cite{jiabocharacterizing} and \cite{jiaboimplications},
I/O wait time means the time spent by CPU waiting for I/O operations to complete.
A high percentage of I/O wait time means that the application has I/O operations frequently,
which further indicates that the application is an I/O intensive workload.
Starting from this point, we describe each application.
\begin{itemize}
\item\textbf{Sort:} Sort simply uses the MapReduce framework to sort records within a directory.
It is a representative I/O intensive application.
\item\textbf{Wordcount:} Wordcount reads text files and counts how often the words occur.
It is a representative CPU intensive application,
accompanied with lighter network and disk load.
\item\textbf{Grep:} The main process in Grep is to find the words provided by users, and count the number of occurrences.
In general, Grep is a CPU intensive application.
\item\textbf{Naive Bayes:} Naive Bayes is a simple probabilistic classifier
which applied the Bayes' theorem with strong (naive) independence assumptions.
We simply select the classification step as our workloads rather than the training step.
The main process is to calculate the probability,
then decide the classification according to the existing model.
It belongs to CPU intensive application.
\item\textbf{SVM:} SVM is a supervised learning model with associated learning algorithms that analyze data and recognize patterns,
used for classification and regression analysis.
SVM is the most complex application in all fives.
The main process of it is vector computation,
and it is a typical CPU intensive application.
\end{itemize}

\subsection{Metrics}
We call for some metrics which can be directly perceived, what's more,
they can be compared and gotten easily.
Also, these metrics should reflect the integrated  processing capacity of big data systems.
We evaluate the performance of a big data system from two aspects \--- processing capacity and energy consumption,
and the corresponding metrics are the DPS and DPJ \cite{luo2012cloudrank}.
DPS is the abbreviation of \textit{Data Processed Per Second}, and DPJ refers to \textit{Data Processed Per Joule}.
The calculation formulas of these two metrics are showed below.
\begin{equation} \label{eq:dps}
  DPS =\frac{Data ~ Input ~ Size}{Run ~Time}
\end{equation}
\begin{equation} \label{eq:dpj}
  DPJ =\frac{Data ~ Input ~ Size}{Energy ~Consumption}
\end{equation}

\textit{Data Input Size} is the data volume of the input workload,
\textit{Run Time} means the execution time,
and \textit{Energy consumption} is the total energy cost of running this workload. We use a power meter to measure energy consumption.
These two metrics aren't inclined to certain components such as the computing power of the CPU or bandwidth of I/O, they place emphasis on estimating the processing capacity of the whole system.
\section{Results and Analysis}
In this part, we will report the experimental data collected in general,
then analyze the performance difference of these big data systems through comparison.
\subsection{General Description}
\begin{figure*}
\begin{minipage}[t]{0.5\linewidth}
\centering
\includegraphics[width=\textwidth]{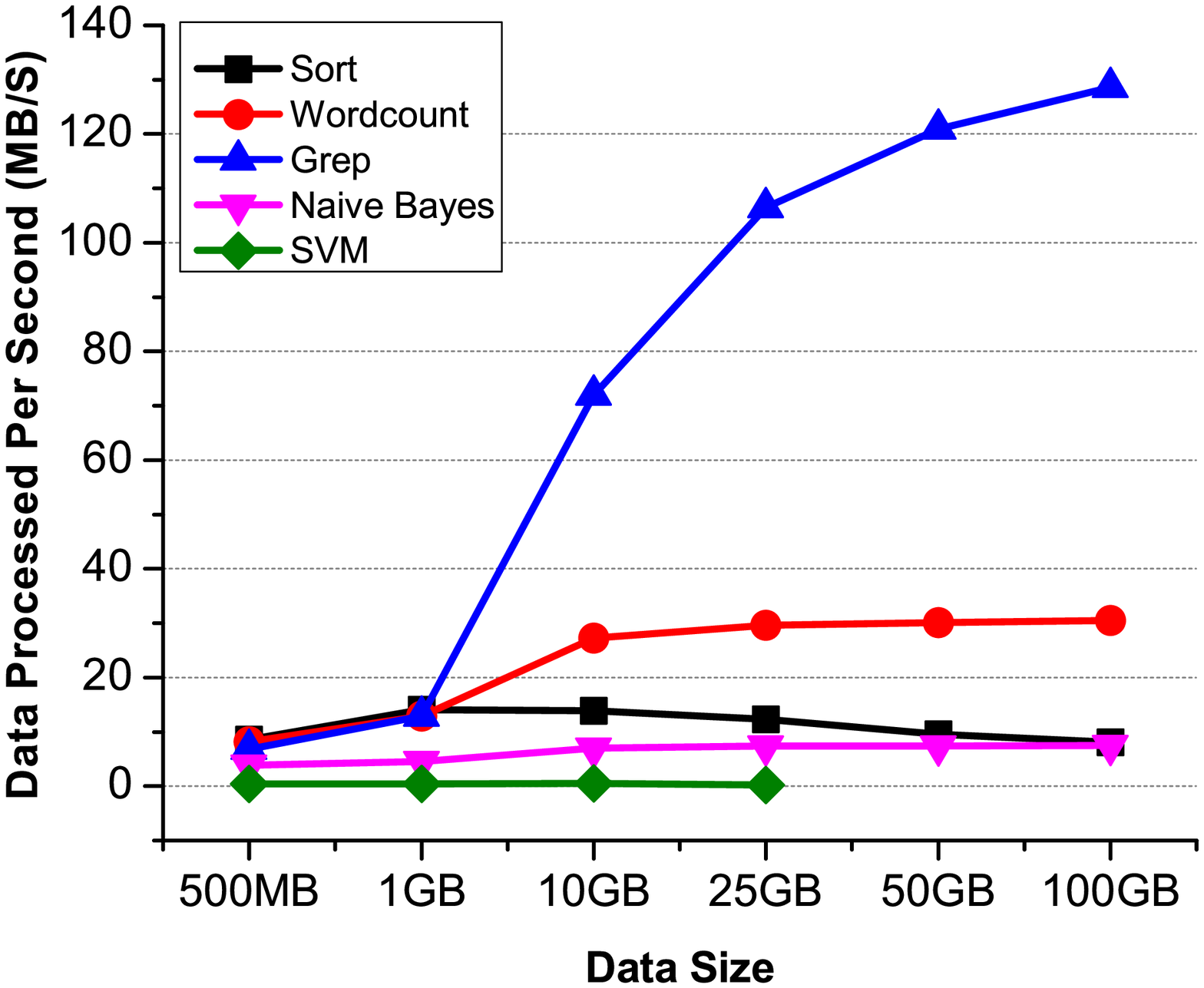}
\caption{The DPS of Xeon Platform}
\label{figure:xeondps}
\end{minipage}%
\begin{minipage}[t]{0.5\linewidth}
\centering
\includegraphics[width=\textwidth]{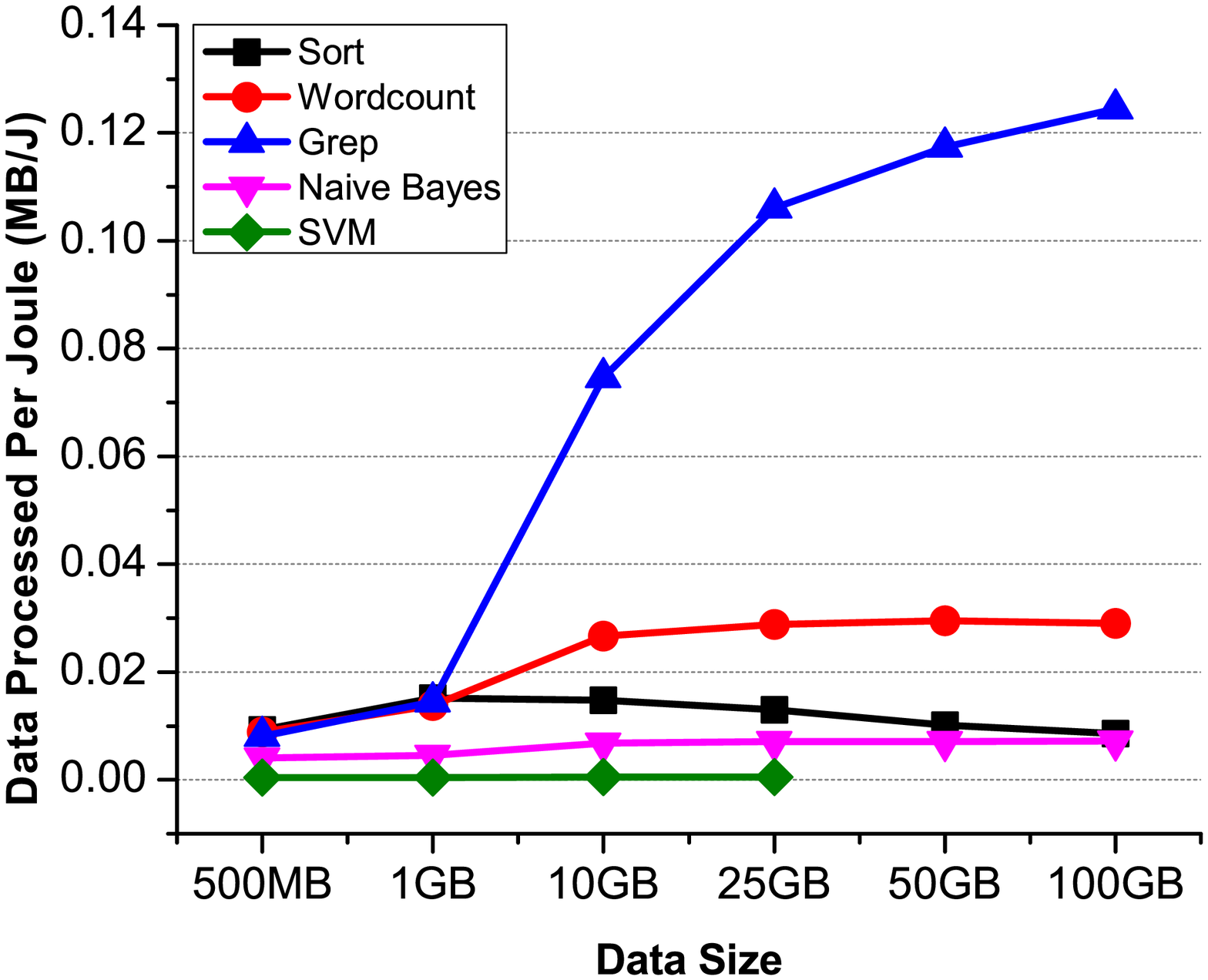}
\caption{The DPJ of Xeon Platform}
\label{figure:xeondpj}
\end{minipage}
\end{figure*}

\begin{figure*}
\begin{minipage}[t]{0.5\linewidth}
\centering
\includegraphics[width=\textwidth]{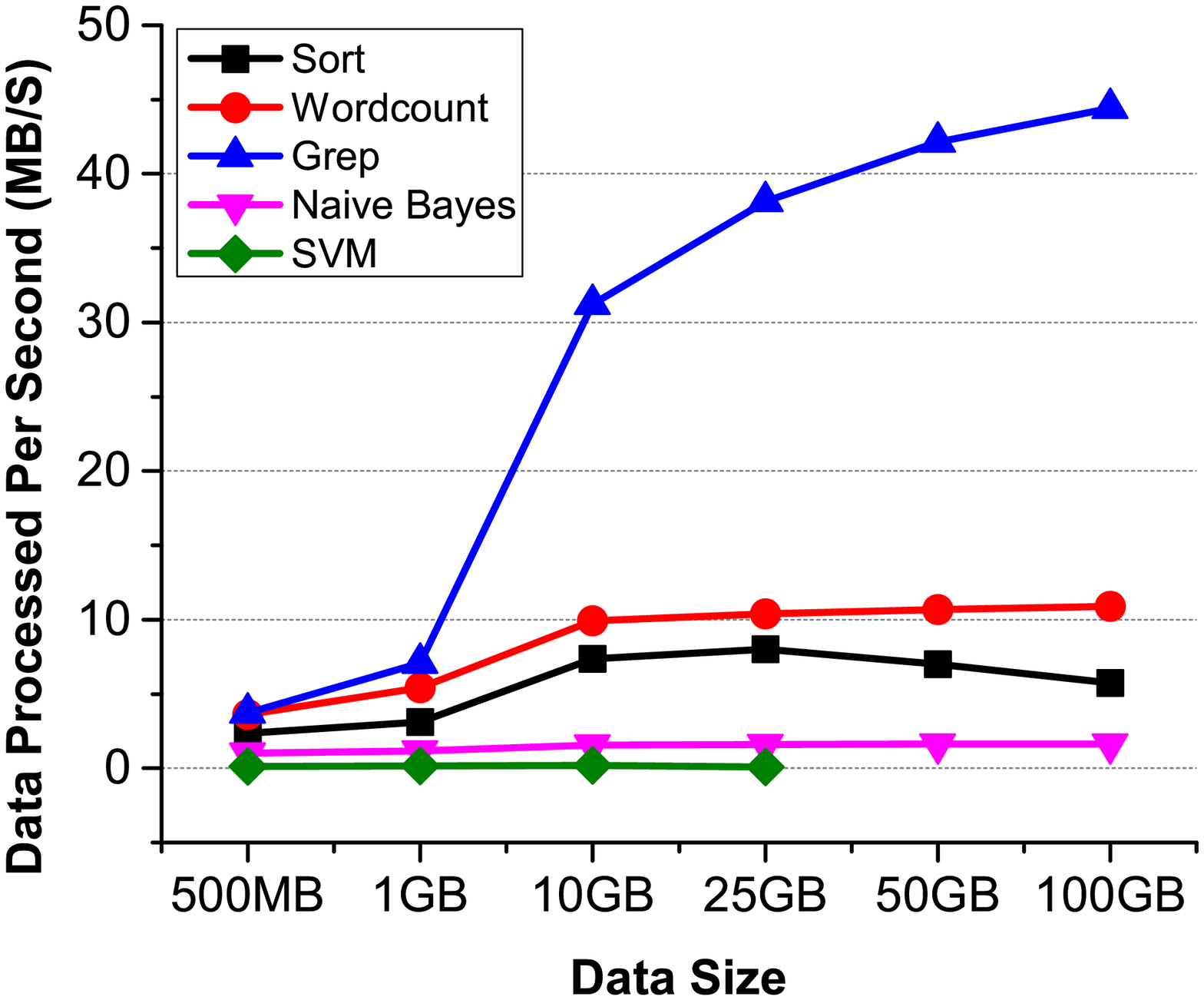}
\caption{The DPS of Atom Platform}
\label{figure:atomdps}
\end{minipage}%
\begin{minipage}[t]{0.5\linewidth}
\centering
\includegraphics[width=\textwidth]{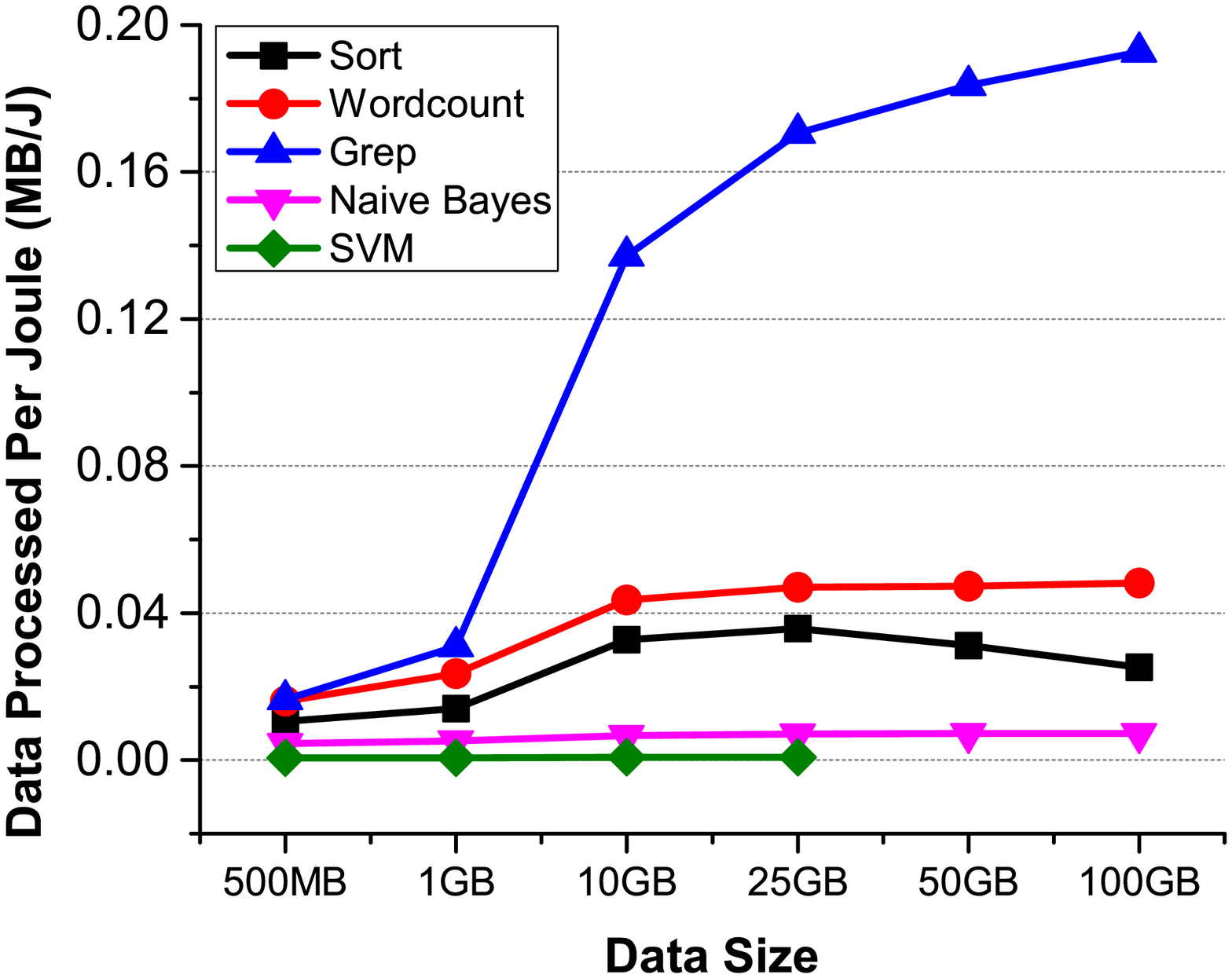}
\caption{The DPJ of Atom Platform}
\label{figure:atomdpj}
\end{minipage}
\end{figure*}

\begin{figure*}
\begin{minipage}[t]{0.5\linewidth}
\centering
\includegraphics[width=\textwidth]{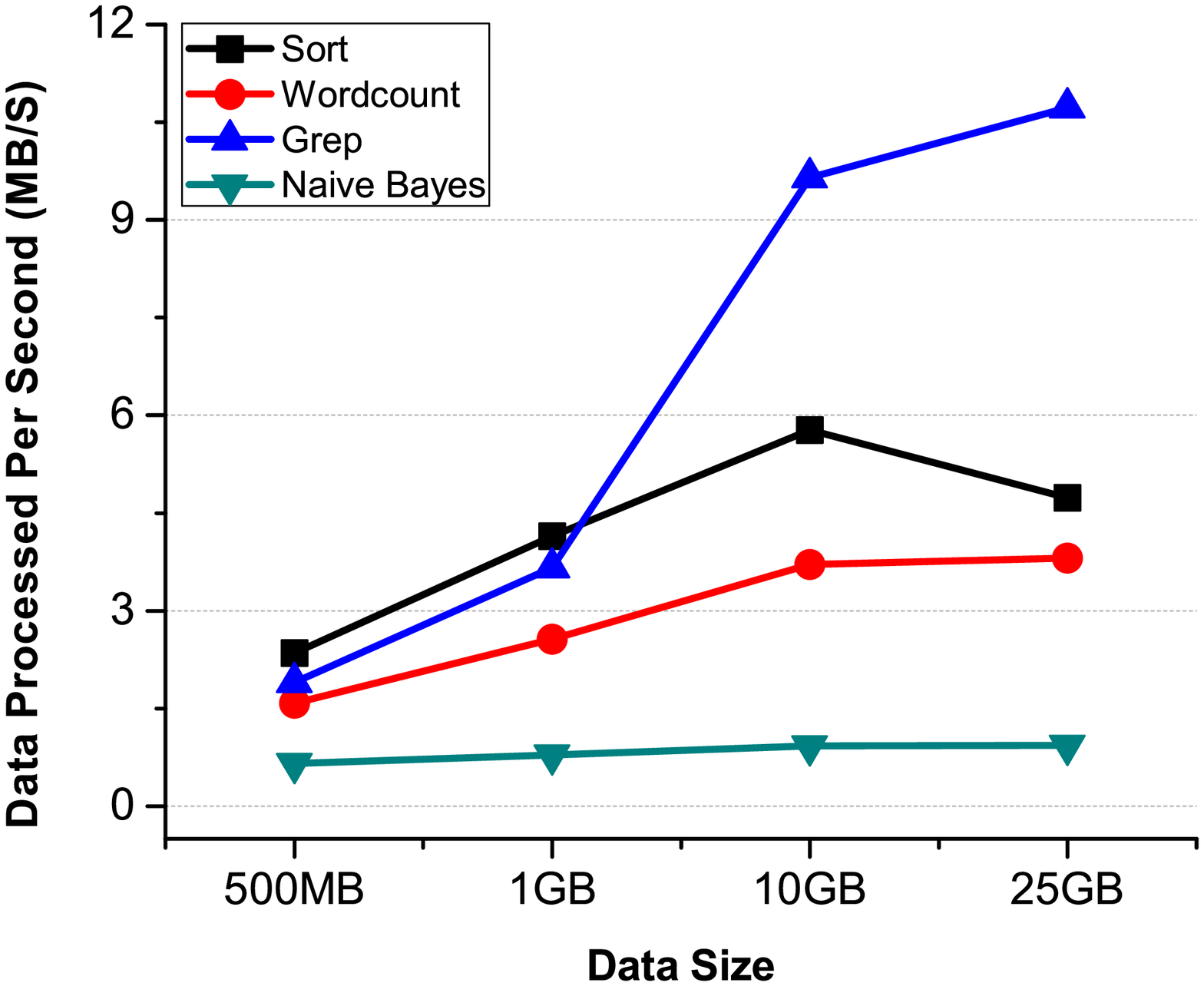}
\caption{The DPS of Tilera Platform}
\label{figure:tildps}
\end{minipage}%
\begin{minipage}[t]{0.5\linewidth}
\centering
\includegraphics[width=\textwidth]{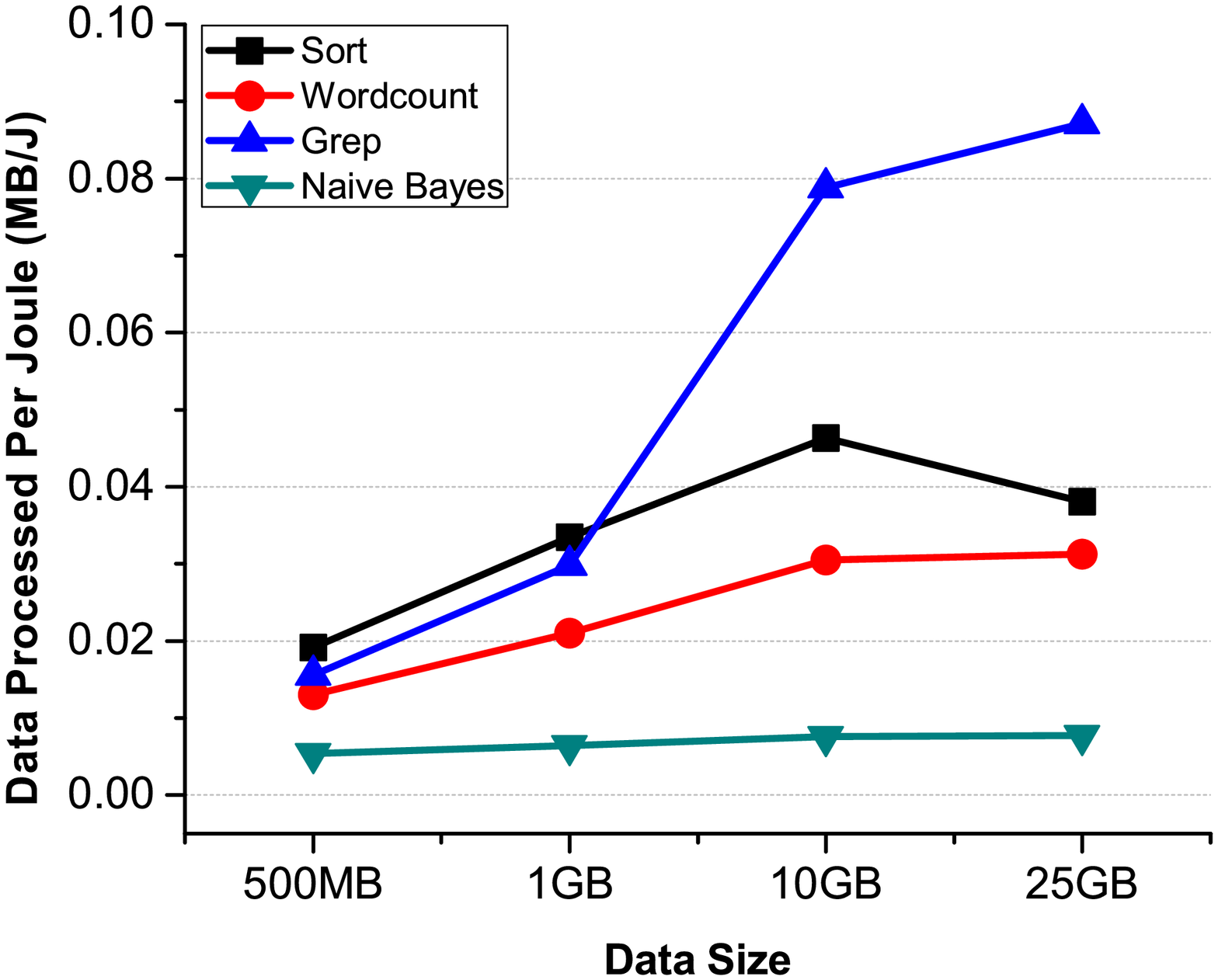}
\caption{The DPJ of Tilera Platform}
\label{figure:tildpj}
\end{minipage}
\end{figure*}
To decide proper data set to drive these systems,
we initially run the experiments using data sets with 500MB and 100GB,
to get approximate processing capacity of system.
During the experiments, we find that SVM is the most time-consuming application,
and the variation of its DPS/DPJ is very small,
so we determine to run the SVM with maximum 25G on Xeon and Atom.
In order to guarantee the stability of our results,
for every experiment, we run at least two times,
taking the average value to eliminate bias caused by uncertainty factors as possible.
Normally, the run time of the same workload are similar, however,
there are existing some abnormal results,
such as one execution time is much higher than the others.
These may be caused by Hadoop abnormal execution --
having too many failed tasks, or the interference from the operating system self-check.
For these kinds of results, we don't take it into our mean value calculation.

Fig. \ref{figure:xeondps} to Fig. \ref{figure:tildpj} display the final DPS and DPJ of three big data systems.
From these figures, we have the following observations.
On three hardware platforms, Grep has the highest DPS/DPJ value, then Wordcount, Sort, and Naive Bayes.
SVM has the smallest ones.
Secondly, for most applications,
along with an increase in the amount of input data,
the DPS/DPJ first rise, and then keep stable, thereamong,
the curve of Sort appears relatively obvious downtrend when the data volume arrive to a certain value,
and the variation of SVM curve is minimal.
Last, the profile of the DPS and the DPJ curve of one system are similar.
\begin{table}
\caption{Details of  Different Algorithms}\label{alogrithm}
\center
\begin{tabular}{|c|c|c|}
  \hline
  Application & Time Complexity & Characteristics  \\ \hline
  Sort & $O(n \times log_2n)$ & Integer comparison  \\ \hline
  WordCount & O(n) & Integer comparison and calculation  \\ \hline
  Grep & O(n) & String comparison  \\ \hline
  Naive Bayes & $O(m \times n)$ & Floating-point computation \\ \hline
  SVM & $O(n^3)$ & Floating-point computation \\
  \hline
\end{tabular}
\end{table}
To the first observation, we find that the performance indicated by the DPS/DPJ
and the growth rate of the curves are closely linked with the complexity of the application.
TABLE \ref{alogrithm} records the computation complexity of the five applications.
Basically, the easier the application is, the higher its DPS/DPJ are.
To observation two, Sort is a typical I/O-intensive application,
and its processing capability trend is mostly impacted by the I/O operations.
I/O wait time directly influence the Sort's execution time.
The more time that spends on I/O wait will reduce the processing capacity.
While, for Wordcount and Grep, they are not I/O-intensive applications,
when the system's resource is fully used, the processing capability hold stable.
For the complex applications like SVM, Naive Bayes,
the changes of the DPS/DPJ is not obvious with the increasing of data volume,
this may be caused by the usage of CPU.
For this kind of applications, CPU is always treated as bottleneck,
the computational resources are fully used when even running small dataset.
To the last observation, by observing the formulas \ref{eq:dps} and \ref{eq:dpj}, we know that,
for the DPS/DPJ, execution time is the most influential factor.
This explains why the curve of the DPS and the DPJ are similar.
Then we dissect the performance difference through comparison.
\subsection{The Comparison of Xeon and Atom}
Both of Xeon and Atom platforms consist of one master and seven slaves,
and the hardware thread number of two systems are the same
(the details of hardware can be found in TABLE \ref{xeon} and TABLE \ref{atom}).

Basic trends of two systems are showed in Fig \ref{figure:xeondps} to \ref{figure:atomdpj}.
\begin{figure*}
\centering
\includegraphics[width=7.5in,height=3.5in]{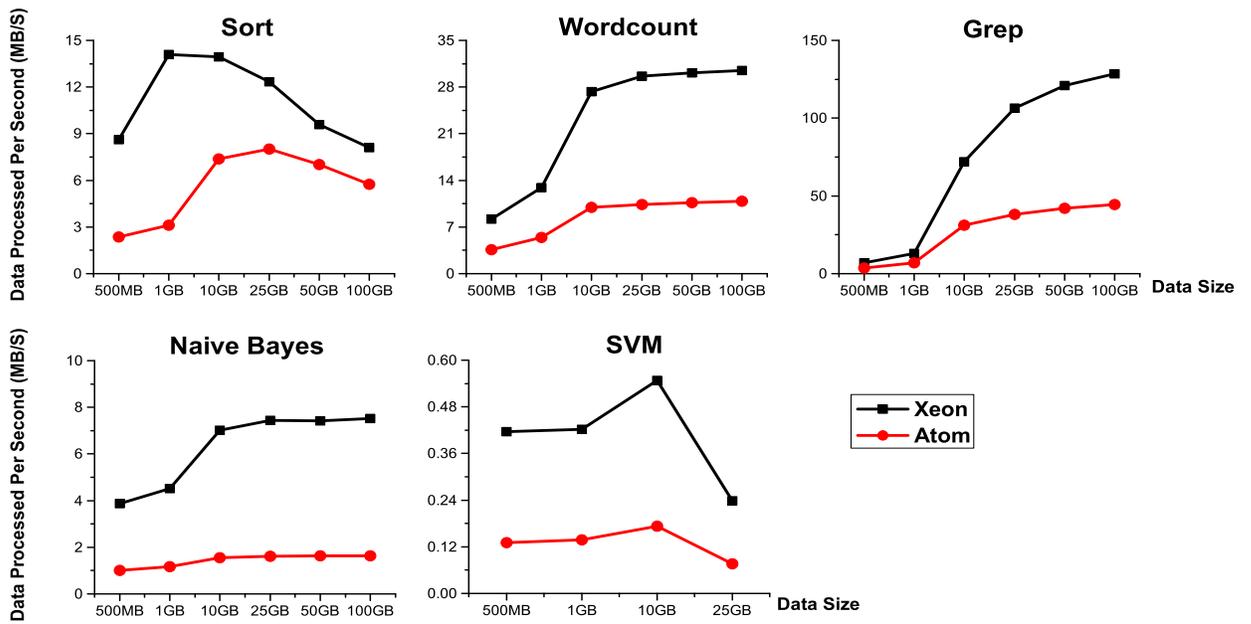}
\caption{The Comparison of DPS between Xeon and Atom }
\label{figure:xadps}
\end{figure*}
\begin{figure*}
\centering
\includegraphics[width=7.5in,height=3.5in]{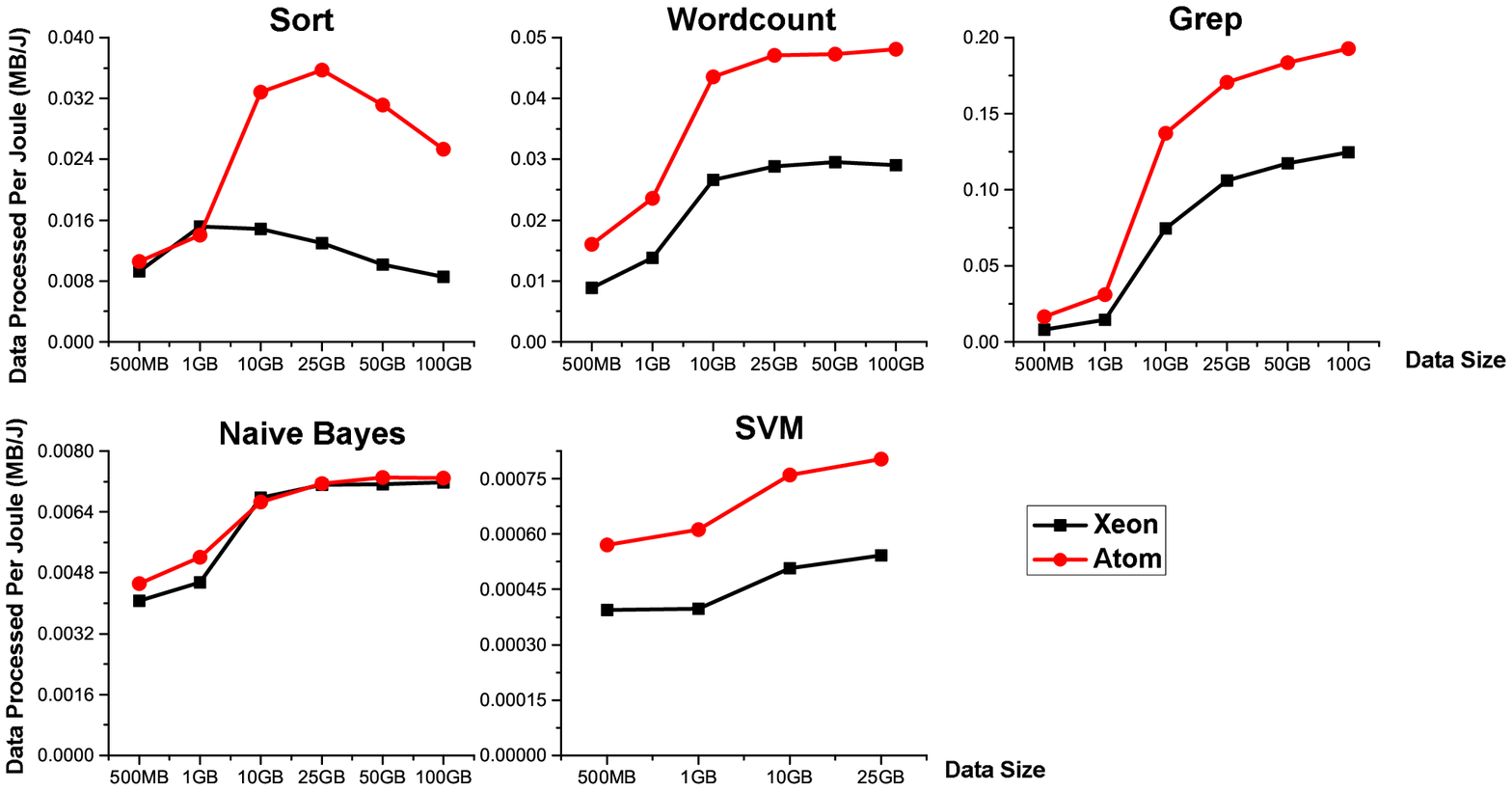}
\caption{The Comparison of DPJ between Xeon and Atom }
\label{figure:xadpj}
\end{figure*}\begin{figure*}
\centering
\includegraphics[width=7.5in,height=2.1in]{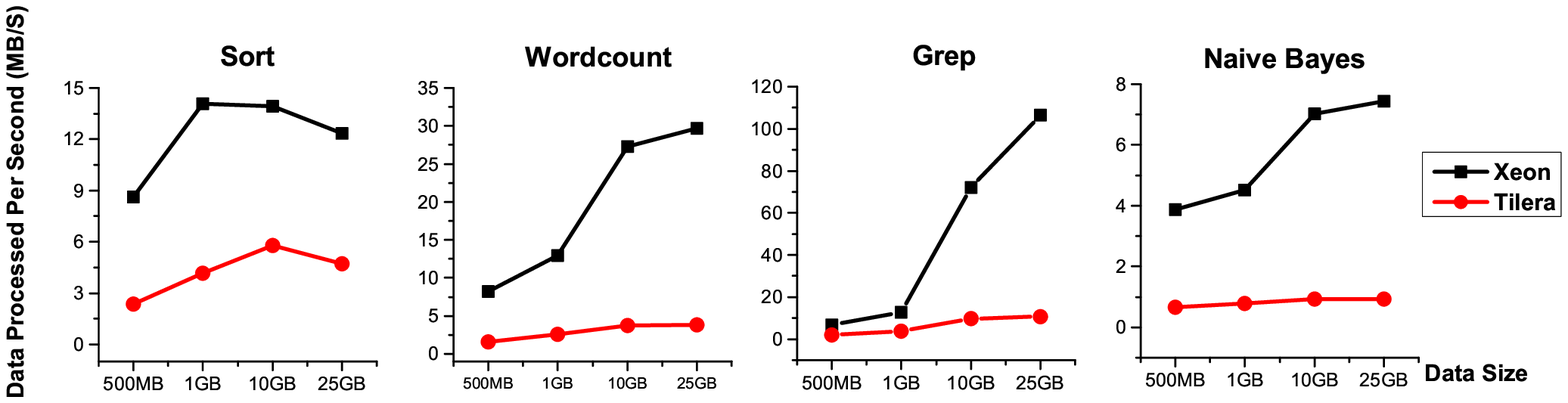}
\caption{The Comparison of DPS between Xeon and Tilera }
\label{figure:xtdps}
\end{figure*}
\begin{figure*}
\centering
\includegraphics[width=7.5in,height=2.1in]{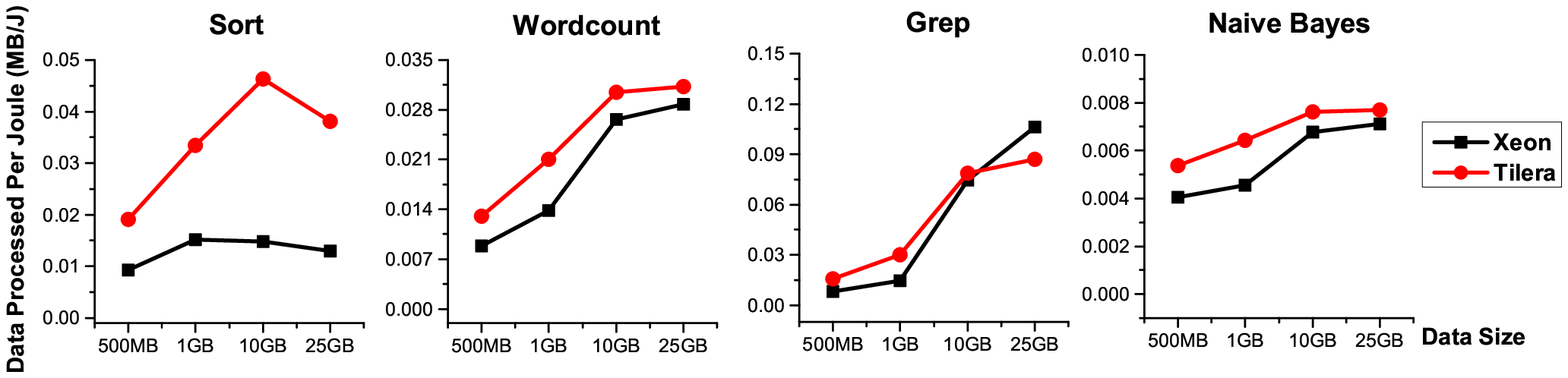}
\caption{The Comparison of DPJ between Xeon and Tilera }
\label{figure:xtdpj}
\end{figure*}
Then Fig. \ref{figure:xadps} and Fig. \ref{figure:xadpj} demonstrate
the \textit{DPS} and the \textit{DPJ} comparison of Xeon and Atom.
Obviously, there is a big gap between Xeon and Atom on processing capacity.
The application type and data volume do an enormous influence on the DPS for both platforms.
For Sort, the distance between the DPS of Xeon and Atom comes closer as the volume of data increases,
yet, for Wordcount, Grep, Naive Bayes, it comes farer with the increasing process.
SVM shows relatively stable state.
We can also see this from TABLE \ref{ratioxa}.
TABLE \ref{ratioxa} shows the ratio of Xeon's DPS/DPJ and Atom's.
We can know that the processing capacity of Xeon basically keeps about 2 to 4 times higher than Atom at most of the time.
For Sort, Wordcount, Grep, the ratio changes variably under different data size,
and for Naive Bayes and SVM, the variation of ratio varies in a small range.
The reason why the DPS/DPJ present such trend may like this.
Sort is a I/O intensive application mentioned above.
The increment of data volume will give rises to the stress of I/O,
which weaken the effects from the CPU processing capacity.
To Wordcount, Grep, Naive Bayes,
when using small dataset, CPU resources are not fully utilized,
so the difference between the two DPS is small.
After increasing the amount of data, CPU performance distance is reflected.
The basic frequency of Xeon and Atom core are 1.6 GHz and 1.66 GHz,
and their hardware thread number on per processor is the same.
The performance difference mainly comes from the pipeline structure.
For Xeon, it supports Out-Of-Order execution while Atom doesn't,
and the work speed is optimized through the pipelining design
which leads to better processing capacity.
For SVM, the CPU utilization are high from the beginning to both platforms which differ from the Sort application.
When using 500M data set, the CPU utilization of Sort only 59\% on Atom,
while SVM arrives to 87\%.
So, to the SVM,
the volume of data set doesn't have a big influence to the two systems' processing capacity distance.

\begin{figure}
\centering
\includegraphics[width=3.6in,height=2.5in]{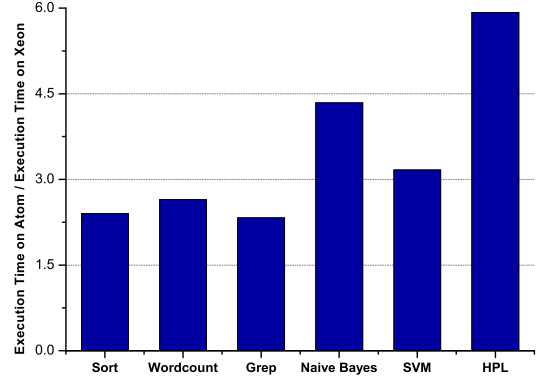}
\caption{The Speedup between Xeon and Atom }
\label{figure:speedup}
\end{figure}

In the meantime, we take the energy consumption into consideration.
Atom is designed primarily for energy conservation. From the Fig. \ref{figure:xadpj},
to Sort, Wordcount, Grep, when the size of data set reaches to a certain value (above 1GB or 10GB for our experimental platforms),
power consumption advantages of Atom are reflected.
This is because of the Xeon system hasn't been fully used under small data volume,
and its energy consumption isn't high in this situation.
To Naive Bayes, Atom does not show the energy strength.
This is caused by too long execution time.
Combined with TABLE \ref{ratioxa}, we see the DPS of SVM on Xeon is more than 3 times higher than Atom's,
and the DPJ of Atom is only 1.5 times higher than Xeon's all the time,
which implies that the Atom is not suitable for such applications either.
To further explain this, we adopt the HPL (High Performance Computing Linpack Benchmark) testing.
We divide the execution time on Atom by the execution time on Xeon to get the speedup.
The Fig. \ref{figure:speedup} shows the result.
From the figure, we learn that the HPL has the highest speedup,
then Naive Bayes and SVM.
The larger value signifies that it will take more time to deal with this application.
For example, it may take 4 times longer if using Atom rather than Xeon, even Xeon cost more energy,
it take less time, the ratio of energy and time still may be lower.
And that is the reason why the Atom doesn't reflect the energy advantage when handling the complex applications.

\begin{table}
\scriptsize
\caption{The ratio between  Xeon DPS/DPJ and Atom DPS/DPJ}\label{xar}
\label{ratioxa}
\center
\begin{tabular}{|c|c|c|c|c|c|c|c|}
\hline
 \multicolumn{2}{|c|}{}& 500MB & 1GB & 10GB & 25GB&50GB&100GB   \\
 \hline
  \multirow{2}{*}{Sort}&DPS & 3.67 & 4.51 & 1.89 & 1.54 & 1.36 & 1.40 \\
  &DPJ & 0.87 & 1.08 & 0.45 & 0.36 & 0.32 & 0.33 \\ \hline
  \multirow{2}{*}{Wordcount}&DPS & 2.27 & 2.38 & 2.74 & 2.84 & 2.82 & 2.79 \\
  &DPJ & 0.55 & 0.58 & 0.61 & 0.61 & 0.62 & 0.60 \\ \hline
  \multirow{2}{*}{Grep}&DPS & 1.83 & 1.82 & 2.30 & 2.79 & 2.87 & 2.89 \\
  &DPJ & 0.48 & 0.46 & 0.54 & 0.62 & 0.63 & 0.64 \\ \hline
  \multirow{2}{*}{Naive Bayes}&DPS & 3.83 & 3.89 & 4.52 & 4.64 & 4.54 & 4.58 \\
  &DPJ & 0.89 & 0.87 & 1.01 & 0.99 & 0.97 & 0.90 \\ \hline
  \multirow{2}{*}{SVM}&DPS & 3.19 & 3.06 & 3.17 & 3.14 &\multirow{2}{*}{\backslashbox{}{}} &\multirow{2}{*}{\backslashbox{}{}} \\
  &DPJ & 0.69 & 0.64 & 0.66 & 0.67 & &\\ \hline
\end{tabular}
\end{table}

\begin{table}
\caption{The ratio between Xeon DPS/DPJ and Tilera DPS/DPJ}\label{xtr}
\label{ratioxt}
\center
\begin{tabular}{|c|c|c|c|c|c|}
\hline
 \multicolumn{2}{|c|}{}& 500MB & 1GB & 10GB & 25GB   \\
 \hline
 \multirow{2}{*}{Sort}&DPS&3.67&3.39&2.41&2.60\\
 &DPJ&0.48&0.45&0.31&0.34\\
\hline
\multirow{2}{*}{Wordcount}&DPS& 5.19& 5.04 & 7.35 & 7.78  \\
 &DPJ & 0.67 & 0.65 & 0.87 & 0.92  \\
\hline
\multirow{2}{*}{Grep}&DPS & 3.60 & 3.52 & 7.45&9.93\\
 &DPJ& 0.51 & 0.48 & 0.94 & 1.21\\
\hline
\multirow{2}{*}{Naive Bayes}&DPS& 5.91 & 5.78 & 7.59& 7.94\\
 &DPJ& 0.75 & 0.70 & 0.89 &0.92\\
\hline
\end{tabular}
\end{table}
\subsection{The Comparison of Xeon and Tilera}
The Tilera processor we use has 36 tiles (cores).
We close 8 tiles (cores) of it to guarantee the core number of two systems to be the same.
Fig. \ref{figure:tildps} and Fig. \ref{figure:tildpj} are the DPS and the DPJ of Tilera system.
By observing the results, we find that on Xeon and Atom, the DPS of Wordcount is larger than Sort,
while on Tilera, the situation is just the opposite.
As we mentioned before, the Sort is a typical I/O intensive application,
so the cache size will make a big difference on its execution time.
TilePro36 integrates 36 tiles (we use 28 tiles), and supports virtual L3 cache mentioned in Section 2. If misses in the L2 cache on a certain tile are satisfied by caches in other tiles,
it will get this data through 2 dimensional on-chip mesh network, otherwise,
Tilera will fetch data from external memory and deliver it to the requesting core \cite{tilerawhitepaper}.
This mechanism gives Tilera more flexible cache access strategies, makes it have more buffer space when processing data, then makes Tilera more suitable for I/O intensive application, like Sort.
Wordcount is a CPU intensive application,
and Tilera is weak on CPU processing capacity.
This may cause the difference comparing with Xeon and Atom.

From the Fig. \ref{figure:xtdps} and Fig. \ref{figure:xtdpj},
we obtain that the processing capacity of Xeon is better than Tilera,
while Tilera has better energy performance.
Nevertheless, except Sort, the DPS of Xeon is around 6 times on average than the DPS of Tilera,
and the DPJ of Tilera is slightly more than one time than that of Xeon.
TABLE \ref{ratioxt} shows the ratio of Xeon's DPS/DPJ and Tilera's.
This implies that the Tilera is more appropriate for I/O intensive applications rather than CPU intensive ones.
Tilera combines the low-power consumption of slower clock speeds with the increased
throughput of many independent cores \cite{bell2008tile64},
which can be used to illustrate the Tilera's processing features.

Above all, we can see that the Xeon cluster has definitely processing advantage in three systems.
We can also deduce that the application behaviors affect the use of system resource,
which lead to the different performance when dealing with diverse workloads.
From the TABLE \ref{alogrithm},
we can see the main operation of SVM and Naive Bayes are all floating-point computation.
Atom which don't support \textit{OoO} (Out of Order) execution and the Tilera which has no floating computing component are hard to run such applications,
even the Tilera have better mechanism for core connecting which may reduce the time costing on the machines communication.
\section{Conclusion}
In this paper, we have evaluated three big data systems.
Based on the above analysis,
we have concluded that different hardware have their own processing features.
A big data system consisted of specific hardware has various performance
when dealing with types of applications and data sets of different volume.
In some cases, a big data system may have both better processing capacity and energy consumption.
More specifically, through our experimental results,
we have shown Xeon generally has better processing capacity accompanied with high energy consumption,
especially to some light scale-out applications like Sort, Wordcount, Grep.
For some complex applications like Naive Bayes, SVM,
the weaker processing capability of processor causes long execution time.
Even the power of them is lower than Xeon, the total energy consumption is still high,
which further illustrate that the energy performance is relative to the application type.
For energy consumption,
Tilera exerts a energy advantage on processing I/O intensive application like Sort,
while Atom is more excellent in processing simple CPU intensive application like Wordcount and Grep.
When architects plan to construct big data systems or choose fundamental components for a system,
they should not only concentrate on the processing capacity and energy consumption of hardware itself,
but also need to regard the application type, data volume and the complexity of the application that the system will handle with.

\section*{Acknowledgment}
We are grateful to our mentors for the many informative discussions that guided this work and
the anonymous reviewers, who provided excellent feedback.
This work is supported by the Chinese 973 project (Grant No.2011CB302502),
the NSFC project (Grant No.60933003, 61202075), and the BNSF project (Grant No.4133081).

\bibliographystyle{IEEEtran}
\bibliography{tex}

\end{document}